\documentclass[showpacs,amsmath,amssymb,aps,pra,
10pt,reprint,superscriptaddress,showkeys]{revtex4-1}
\usepackage{graphicx}
\usepackage{bm}
\usepackage[breaklinks=true,colorlinks=true,linkcolor=blue,urlcolor=blue,citecolor=blue]{hyperref}
\usepackage{graphics}
\usepackage{dcolumn}
\usepackage{amsmath,amssymb}
\usepackage{mathdots}
\usepackage{natbib}
\usepackage{soul,color}
\usepackage{epstopdf}
\usepackage[note-name]{notes2bib}
\usepackage[utf8]{inputenc}
\usepackage[T1]{fontenc}
\usepackage{mathptmx} 

\begin{document}

\title{\Large Cherenkov radiation of spin waves by ultra-fast moving magnetic flux quanta}

\author{O. V.~Dobrovolskiy}
    \email{oleksandr.dobrovolskiy@univie.ac.at}
    \affiliation{Faculty of Physics, University of Vienna, 1090 Vienna, Austria}
\author{Q. Wang}
    \affiliation{Faculty of Physics, University of Vienna, 1090 Vienna, Austria}
\author{D.\,Yu.\,Vodolazov}
     \affiliation{Independent Researcher}
\author{B. Budinska}
    \affiliation{Faculty of Physics, University of Vienna, 1090 Vienna, Austria}
\author{S. Knauer}
    \affiliation{Faculty of Physics, University of Vienna, 1090 Vienna, Austria}
\author{R. Sachser}
    \affiliation{Physikalisches Institut, Goethe University, 60438 Frankfurt am Main, Germany}
\author{M. Huth}
    \affiliation{Physikalisches Institut, Goethe University, 60438 Frankfurt am Main, Germany}
\author{A.\,I. Buzdin}
    \affiliation{Universit\'{e} Bordeaux, CNRS, LOMA, UMR 5798, F-33405 Talence, France}

\begin{abstract}
Despite theoretical predictions for a Cherenkov-type radiation of spin waves (magnons) by various propagating magnetic perturbations, fast-enough moving magnetic field stimuli have not been available so far.
Here, we experimentally realize the Cherenkov radiation of spin waves in a Co-Fe magnonic conduit by fast-moving ($>$1\,km/s) magnetic flux quanta (Abrikosov vortices) in an adjacent Nb-C superconducting strip.
The radiation is evidenced by the microwave detection of spin waves propagating a distance of 2\,$\mu$m from the superconductor and it is accompanied by a magnon Shapiro step in its current-voltage curve.
The spin-wave excitation is unidirectional and monochromatic, with sub-40\,nm wavelengths determined by the period of the vortex lattice.
The phase-locking of the vortex lattice with the excited spin wave limits the vortex velocity and reduces the dissipation in the superconductor.
\end{abstract}
\maketitle

Back in 1887, Lord Kelvin was fascinated by complex wave patterns generated behind his boat on the surface of water\,\cite{Tho87prs}. The generated patterns of waves resembled the V letter, with the boat being ahead of the waves and giving rise to the wake of the apex. Ever since the formation of wakes by a perturbation moving faster than the propagation speed of the waves it creates has been established as a universal phenomenon, and it now plays a significant role across various disciplines. Thus, counterparts of the Kelvin wake in hydrodynamics are the well-known sonic boom in acoustics and Cherenkov radiation in electrodynamics. The Cherenkov effect describes the spontaneous emission of photons in a medium, which occurs when a charged particle moves at a velocity faster than the phase velocity of light in that medium\,\cite{Che34dan,Che58npp}. Together with the Doppler effect, the Cherenkov effect constitutes the branch of fundamental physics describing the radiation of uniformly moving sources\,\cite{Gin96phu}. Awarded with the Nobel Prize in 1958\,\cite{Che58npp}, the Cherenkov effect finds applications in detectors in particle physics\,\cite{Gru00acp} and cosmology\,\cite{Vri11prl}, and it plays a critical role in photonics\,\cite{Luo03sci,Riv20nrp}, electromagnetics\,\cite{Dua17nac}, biomedicine\,\cite{Gri18nbe}, and across various domains of solid-state physics\,\cite{Gol00prb,Shk18prb,She19prb,Gen15nan,Xia16nsr,Yan11apl,Yan13prb,Bul05prl,She11prl,Bes14prb}.

Among the various types of waves in condensed matter systems, spin waves -- the Goldstone modes of spin systems\,\cite{Blo30zph,Gur96boo} -- represent an essential realm of waves in magnetic materials. Nowadays, spin waves and their quanta -- magnons -- are at the heart of magnonics\,\cite{Kru10jpd,Dem13boo} which has emerged as one of the most rapidly developing research domains of modern magnetism and spintronics\,\cite{Die20nel,Wan20nel,Bar21pcm}. Generation of spin waves by moving magnetic sources via a Cherenkov-type mechanism has been predicted in numerous theoretical works\,\cite{Bou90prl,Bar94inb,Yan11apl,Yan13prb,Bul05prl,She11prl,Bes14prb}. Among various candidate sources to perturb the magnetic moments, moving magnetic monopoles\,\cite{Dat84pla,Kol98etp}, domain walls\,\cite{Bou90prl,Bar94inb,Yan11apl,Yan13prb} and Abrikosov vortices (fluxons)\,\cite{Bul05prl,She11prl,Bes14prb} were theoretically suggested. However, despite the theoretical predictions, fast-enough moving magnetic sources have not been available so far. At the required high velocities of spin waves (few km/s in ferromagnet-based devices), domain walls collapse because of the Walker breakdown\,\cite{Sch74jap} while the lack of long-range order in vortex arrays\,\cite{Emb17nac} makes in-phase generation of spin waves hardly feasible for the majority of superconductor-based systems. We note that domain walls can move significantly faster in ferrimagnets \cite{Car20sci}, antiferromagnets and ferromagnetic nanotubes. However, the immunity of antiferromagnets to magnetic fields presents notorious difficulties in manipulating domain walls, ferrimagnets require to operate in vicinity of the angular momentum compensation temperature \cite{Kim17nam} while high-quality round ferromagnetic nanotubes with sufficiently low damping remain inaccessible so far \,\cite{Kor20arx}.

Recently, we observed a strong magnon-fluxon interaction in a Nb/Py superconductor/ferro\-magnet heterostructure and demonstrated Doppler shifts in the frequency spectra of spin waves scattered on a moving vortex lattice\,\cite{Dob19nph}. However, the sub-km/s maximal vortex velocities in that Nb/Py heterostructure were not high enough for the generation of magnons via a Cherenkov-type mechanism\,\cite{Dob19nph}. Very recently, a direct-write Nb-C superconductor with fast relaxation of heated electrons was discovered\,\cite{Por19acs}. The fast heat removal from nonequilibrium electrons in Nb-C allows for ultra-fast vortex motion with up to 15\,km/s vortex velocities\,\cite{Dob20nac}. Here, we experimentally evidence the Cherenkov radiation of spin waves in a Co-Fe magnonic conduit by fast-moving magnetic flux quanta (Abrikosov vortices) in an adjacent Nb-C superconducting strip. We observe the magnon Cherenkov radiation directly, by means of broadband microwave detection of spin waves traveling over a distance of about 2\,$\mu$m through the magnonic conduit. In addition, we monitor the electric voltage across the superconducting strip which exhibits a constant-voltage Shapiro step at the Cherenkov resonance radiation condition. This magnon Shapiro step emerges because of the phase-locking of the vortex lattice with the excited spin wave which limits the vortex velocity and represents a dynamic pinning mechanism for the reduction of dissipation in superconductor-based heterostructures\,\cite{Bul05prl,She11prl,Bes14prb}. We elucidate the experimental observations with the aid of micromagnetic simulations indicating that the Cherenkov resonance condition corresponds to the intersection point of the dispersion curves for the magnon and fluxon subsystems. Because of the periodicity of the vortex lattice, the spin-wave excitation is unidirectional (spin wave propagates in the direction of motion of the vortex lattice) and monochromatic (spin-wave wavelength is equal to the vortex lattice parameter). The sub-40\,nm wavelengths of the detected spin waves are a factor of about two smaller than the shortest wavelengths of propagating spin waves observed experimentally so far\,\cite{Yuh16nac,Liu18nac,Slu19nan,Die19prl,Che20nac,Yuh21phr}.

\begin{figure}[t!]
    \centering
    \includegraphics[width=0.95\linewidth]{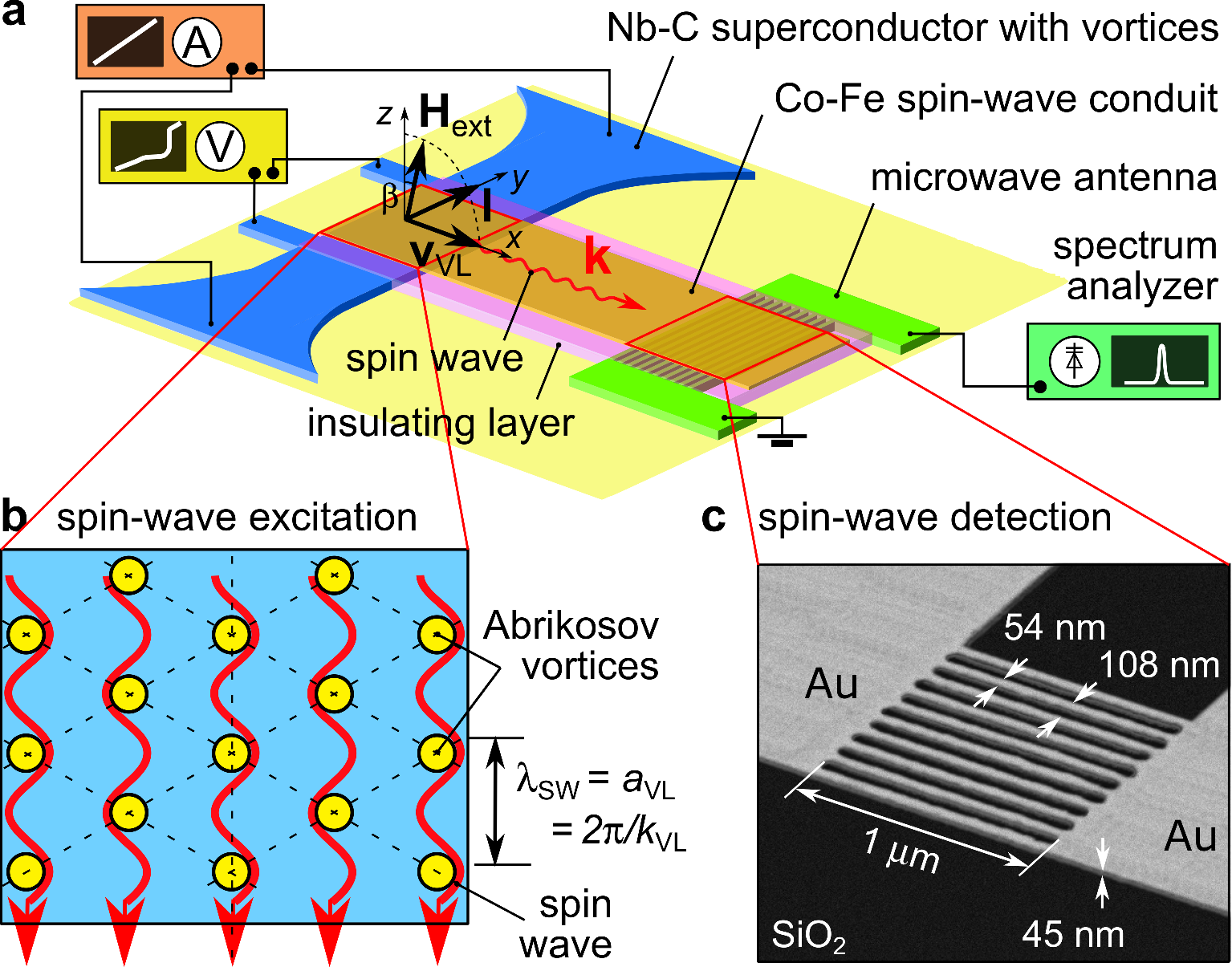}
    \caption{\textbf{Investigated system}.
    \textbf{a} Experimental geometry. The superconductor/ferromagnet hybrid structure consists of a Nb-C superconducting strip stray-field coupled to a Co-Fe magnonic conduit.
    Measurements are taken at 4.2\,K in a magnetic field $H_\mathrm{ext}$ directed at a small angle $\beta\approx5^\circ$ with respect to the $z$ axis in the $xz$ plane.
    \textbf{b}~Schematic of the excitation of spin waves by the moving vortex lattice at the Cherenkov resonance condition.
    \textbf{c} Scanning electron microscopy image of the microwave ladder antenna used for the detection of spin waves with wavelengths $\lambda_\mathrm{SW}\approx36$\,nm (wavevectors $k_\mathrm{SW}\approx175$\,rad/$\mu$m).
    \label{f1}
}
\end{figure}

\textbf{Investigated system and magnon Shapiro steps in its current-voltage curve}.
The investigated system consists of a 45\,nm-thick Nb-C superconducting strip and a 30\,nm-thick Co-Fe ferromagnetic magnonic conduit (Fig.\,\ref{f1}\,\textbf{a}), separated from each other by a 3\,nm-thick insulating layer and interacting via stray fields\,\cite{Gol18afm,Dob19nph}. The measurements are taken at 4.2\,K in the vortex state of Nb-C, below its superconducting transition temperature $T_\mathrm{c} = 5.6$\,K\,\cite{Por19acs}. In an external magnetic field, Nb-C is penetrated by a lattice of Abrikosov vortices (fluxons), each of which carries one quantum of magnetic flux $\Phi_0 = 2.068\times10^{-15}$\,Wb\,\cite{Abr57etp}. The vortices can be imagined as tiny whirls of the supercurrent circulating around cylinders of the material which is in the normal state. The vortex lattice parameter $a_\mathrm{VL} = (2\Phi_0/\sqrt{3}H_\mathrm{ext})^{1/2}$ can be tuned by variation of the external magnetic field value $H_\mathrm{ext}$. The lattice of vortices is characterized by a modulation of the local magnetic field which attains a maximum at the vortex cores\,\cite{Bra95rpp}.

The external magnetic field $H_\mathrm{ext}$ is applied at a small tilt angle $\beta\approx5^\circ$ with respect to the sample normal, in the plane perpendicular to the current direction (Fig.\,\ref{f1}\,\textbf{a}). The applied current $I$ induces a Lorenz force on the lattice of Abrikosov vortices. At sufficiently large transport currents, the vortex lattice moves at velocity $v_\mathrm{VL}$ and induces oscillations of the local magnetic field at a given point in space at the washboard frequency $f_\textrm{WB} = v_\mathrm{VL}/a_\mathrm{VL}$. The applied magnetic field $H_\mathrm{ext}$ is varied between 1.75\,T and 1.95\,T. It is sufficient to magnetize the Co-Fe magnonic conduit to saturation, thus setting the spin-wave propagation to a configuration which is close to the forward volume spin-wave (FVSW) geometry\,\cite{Gur96boo}. The motion of vortices in the superconducting strip triggers a precession of spins in the magnonic conduit\,(Fig.\,\ref{f1}\,\textbf{b}). Once the velocity of the vortex lattice in the superconductor reaches the phase velocity of spin waves in the ferromagnet, the Cherenkov radiation condition is satisfied\,\cite{Bul05prl,She11prl,Bes14prb}. The propagation of the excited spin waves through the magnonic conduit is monitored by a spectrum analyzer connected to a microwave ladder nano-antenna located at a distance of 2\,$\mu$m away from the Nb-C/Co-Fe hybrid region  (Fig.\,\ref{f1}\,\textbf{c}). The supplementary materials contain details on the fabrication and properties of the investigated system.
\begin{figure}[t!]
    \centering
    \includegraphics[width=0.98\linewidth]{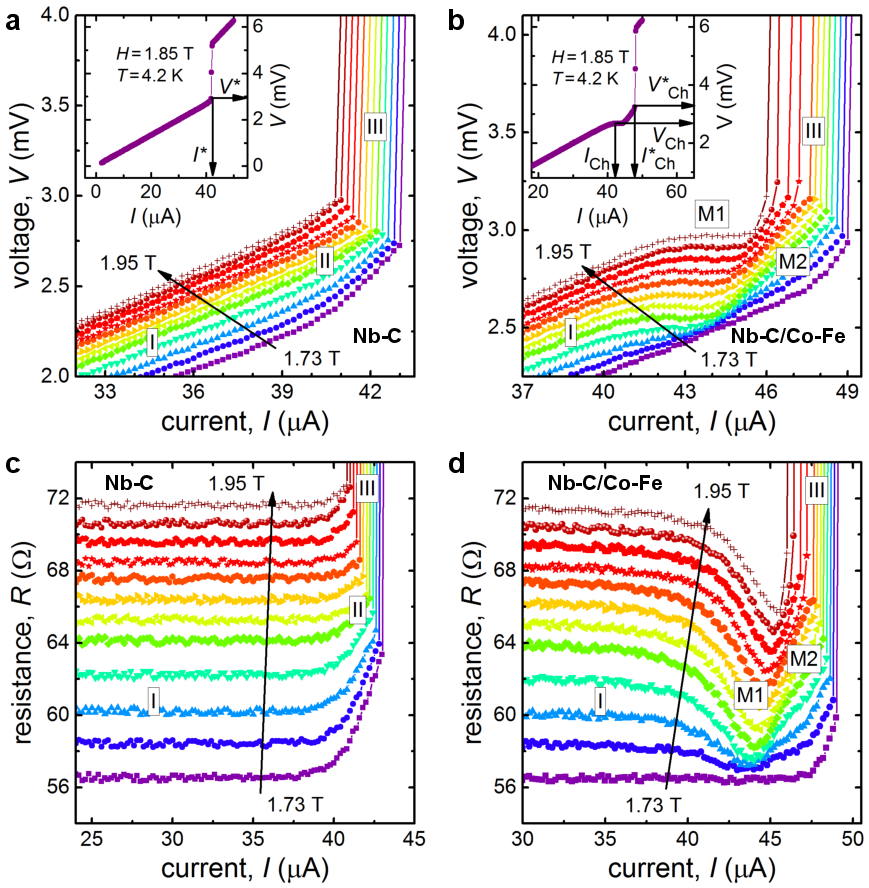}
    \caption{\textbf{Voltage steps upon the Cherenkov radiation of magnons by fluxons}.
    \textbf{a} Nb-C superconductor and \textbf{b} Nb-C/Co-Fe hybrid structure. The magnetic field $H_\mathrm{ext}$ increases in steps of 20\,mT, as indicated. The instability current $I^\ast$, the instability voltage $V^\ast$, and the respective quantities in the regime of Cherenkov generation of magnons by fluxons, $I^\ast_\mathrm{Ch}$ and $V^\ast_\mathrm{Ch}$, are indicated in the insets. $I_\mathrm{Ch}$ and $V_\mathrm{Ch}$ are defined at the center of the voltage steps.
    \textbf{c} and \textbf{d} Evolution of the electrical resistance for the same samples as a function of the transport current.
    \label{f2}
}
\end{figure}

\begin{figure*}[t!]
    \centering
    \includegraphics[width=0.9\linewidth]{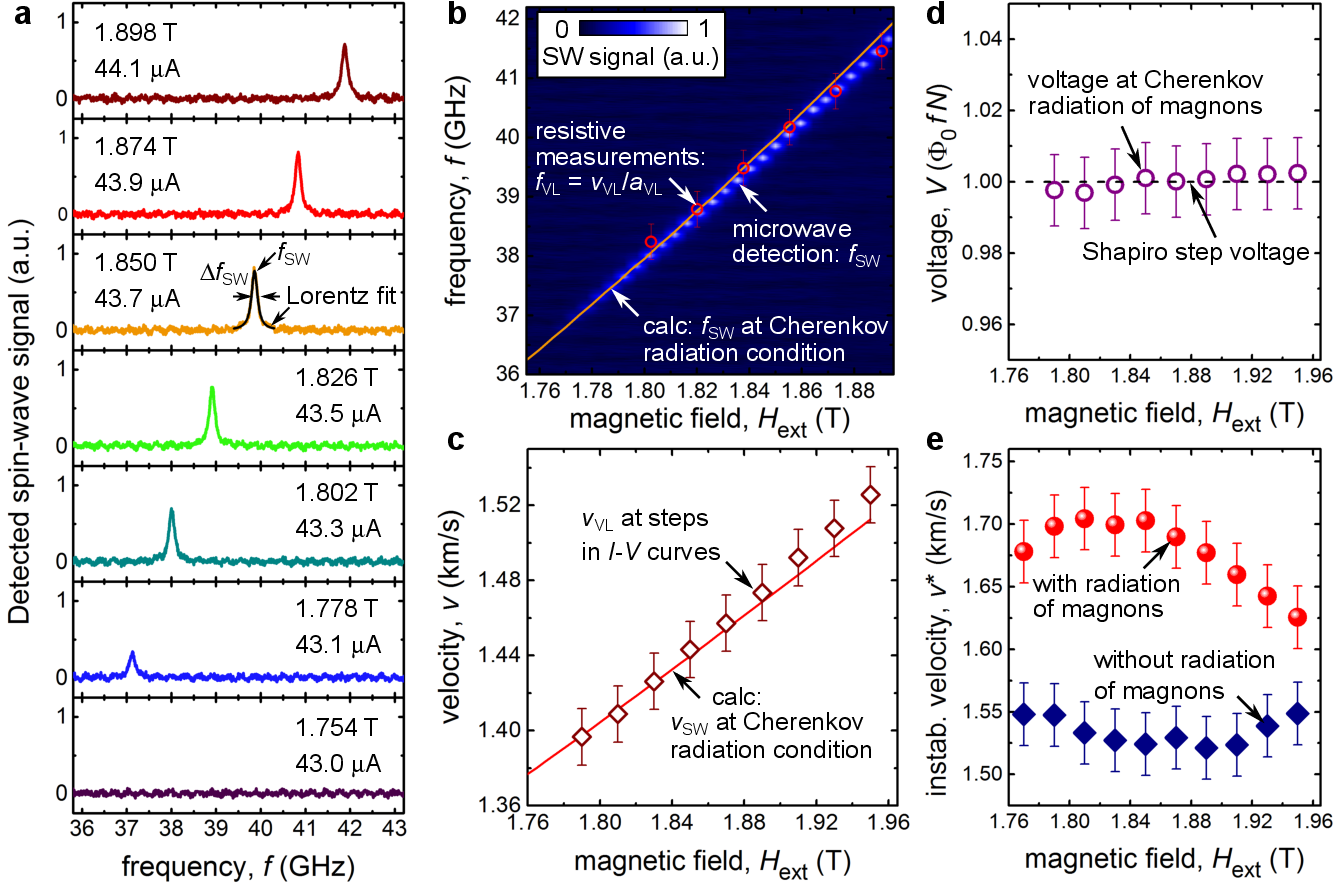}
    \caption{\textbf{Microwave detection of propagating spin waves}.
    \textbf{a} Detected spin-wave spectra for a series of magnetic field and current values.
    \textbf{b}~Detected microwave power versus frequency and magnetic field value.
    \textbf{c}~Vortex velocity $v_\mathrm{VL}$ deduced from the voltage step in the $I$-$V$ curves (symbols) in comparison with the calculated phase velocity of spin waves $v_\mathrm{SW}$ (line) at the Cherenkov resonance condition.
    \textbf{d}~Measured voltage at the steps in the $I$-$V$ curves in comparison with the expected Shapiro step voltage.
    \textbf{e}~Enhancement of the flux-flow instability velocity $v^\ast$ upon Cherenkov radiation of magnons by fluxons.
    \label{f3}
}
\end{figure*}
The viscous motion of vortices in the superconducting strip is associated with a retarded recovery of the superconducting order parameter\,\cite{Bra95rpp}, resulting in an ohmic branch in the current-voltage ($I$-$V$) curve. This behavior is seen for the superconducting strip before the deposition of the magnonic conduit, which serves as a reference measurement in our experiments (Fig.\,\ref{f2}\,\textbf{a}). The $I$-$V$ curves exhibit a nearly linear regime of flux flow (regime I) up to a current of 40\,$\mu$A. At larger currents, the $I$-$V$ curves show a non-linear upturn (regime II) preceding the Larkin-Ovchinnikov instability (regime III)\,\cite{Lar75etp,Dob19pra,Dob20nac}. The flux-flow instability occurs at the instability current $I^\ast$ which is associated with the instability voltage $V^\ast$ (Fig.\,\ref{f2}\,\textbf{a}, inset).

The presence of a magnonic conduit on top of the superconducting strip leads to the appearance of constant-voltage steps in the $I$-$V$ curves (Fig.\,\ref{f2}\,\textbf{b}). The steps occur at voltages $V_\mathrm{Ch}$ (Fig.\,\ref{f2}\,\textbf{b}, inset). The appearance of steps in the $I$-$V$ curves is accompanied by an expansion of the low-resistive regime towards larger currents and an increase of the instability current $I^\ast_\mathrm{Ch}> I^\ast$ and the instability velocity $v^\ast_\mathrm{Ch}> v^\ast$. The electrical resistance $R(I)$ for the reference bare Nb-C strip increases monotonically with increasing current (Fig.\,\ref{f2}\,\textbf{c}). Contrastingly, $R(I)$ for the Nb-C/Co-Fe bilayer exhibits a minimum at the foot of the instability jump (Fig.\,\ref{f2}\,\textbf{d}). We label the constant-voltage regimes with M1 which stands for the Magnonic I regime (Fig.\,\ref{f2}\,\textbf{b} and Fig.\,\ref{f2}\,\textbf{d}). The pronounced nonlinear regime M2 refers to the Magnonic II regime. We will return to the elucidation of these regimes when discussing micromagnetic simulation results.

\textbf{Microwave detection of spin waves and its correlation with resistance measurements}.
For the detection of propagation of the spin waves excited by moving magnetic flux quanta the transport current through the superconductor was tuned to $I_\mathrm{Ch}$, which corresponds to the middle of the voltage steps in the $I$-$V$ curves (Fig. \ref{f2}\,\textbf{b}). The detected microwave signal is peaked at the frequency $f_\mathrm{SW}$ which increases with increase of the external field value (Fig.\,\ref{f3}\,\textbf{a}). We note that no microwave signal was observed for the superconducting strip without magnonic conduit. This means that the detected signal must be related to spin waves propagating through the magnonic conduit rather than being picked up inductively from vortices moving in the superconducting strip\,\cite{Bul05prl,Dob18nac}. In the supplementary materials we demonstrate that the microwave signal disappears upon a current polarity reversal resulting in the co-propagating vortex lattice and spin wave away from the detector antenna, while the voltage steps are maintained. In return, with a further reversal of the magnetic field polarity (that is when both, the current and the external magnetic field are directed oppositely to the directions shown in Fig.\,\ref{f1}\,\textbf{a}) the microwave signal re-appears.

The magnetic-field dependence of the frequency of the detected spin-wave signal, $f_\mathrm{SW}(H_\mathrm{ext})$, is nearly linear (Fig.\,\ref{f3}\,\textbf{b}). The detected peak frequency $f_\mathrm{SW}$ matches, within 5\% accuracy, the washboard frequency of the vortex lattice $f_\mathrm{VL}$. The magnetic field dependence of the vortex velocity, $v_\mathrm{VL}(H_\mathrm{ext})$, deduced from the voltage step in the $I$-$V$ curves, is also nearly linear (Fig.\,\ref{f3}\,\textbf{c}). The observed voltage steps occur at the vortex velocities $v_\mathrm{Ch}$ between 1.38\,km/s and 1.52\,km/s. These velocities are only approximately 50\,m/s smaller than the typical instability velocities $v^\ast$ in the bare Nb-C superconductor, and they are approximately 200\,m/s smaller than $v^\ast_\mathrm{Ch}$ when the superconducting strip is overlaid with a Co-Fe magnonic conduit.
\begin{figure*}[t!]
    \centering
    \includegraphics[width=0.9\linewidth]{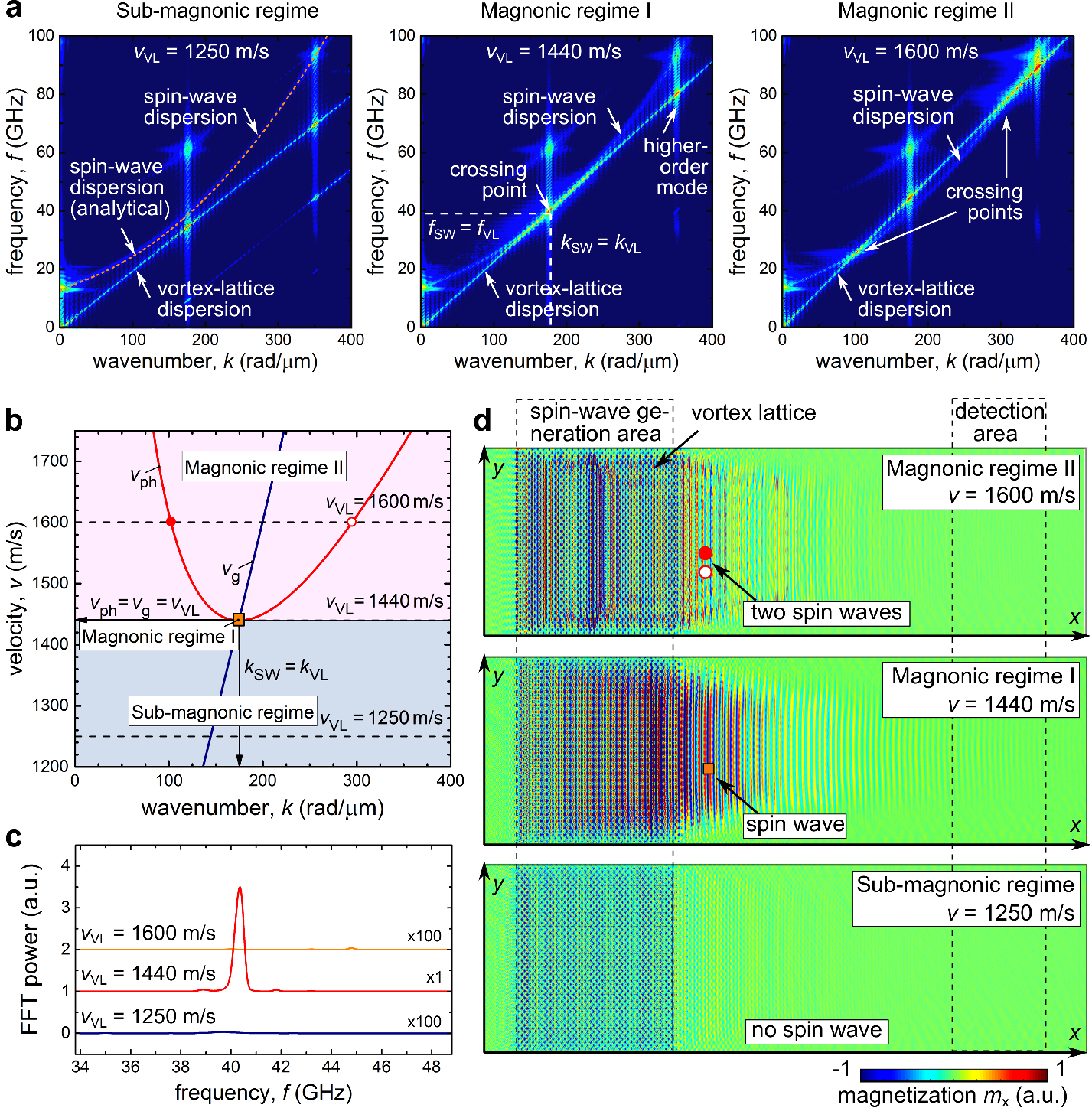}
    \caption{\textbf{Micromagnetic simulations of the Cherenkov radiation of magnons by fluxons}.
    \textbf{a}~Dispersion curves for spin waves at $H_\mathrm{ext}=1.85$\,T and the vortex lattice moving with different velocities.
    \textbf{b} Dependence of the phase ($v_\mathrm{ph}$) and group ($v_\mathrm{g}$) velocities of spin waves on the wavenumber $k$. Magnonic regime I corresponds to a monochromatic and unidirectional spin-wave excitation at the Cherenkov resonance condition $v_\mathrm{VL} = v_\mathrm{SW} = v_\mathrm{ph} = v_\mathrm{g}$ and $k_\mathrm{VL} = k_\mathrm{SW}$.
    \textbf{c} Spin-wave spectra in the detection area with multiplication factors, as indicated.
    \textbf{d} Propagation of spin waves in the magnonic conduit.
    \label{f4}
}
\end{figure*}

A remarkable correlation is found between the step voltage $V_\mathrm{Ch}$ in the $I$-$V$ curves (Fig.\,\ref{f2}\,\textbf{b}) and the peak frequencies $f_\mathrm{SW}$ in the microwave detection (Fig.\,\ref{f3}\,\textbf{a}). The magnetic field dependence of the normalized step voltage expressed in units of ($\Phi_0 f_\mathrm{SW} N$), where $N$ is the number of vortices between the voltage leads, reveals that the step voltage is constant and equal to the product of the spin-wave frequency $f_\mathrm{SW}$ with the number of vortices between the voltage leads and the magnetic flux quantum $\Phi_0$ (Fig.\,\ref{f3}\,\textbf{d}). This finding reveals the fundamental nature of the voltage step associated with the Cherenkov radiation of magnons by fluxons: In the considered system, the Cherenkov radiation (threshold) velocity corresponds to the Shapiro step \cite{Sha63prl,Fio71prl}. The appearance of Shapiro steps is a generic feature of systems where an object, moving in a periodic potential, is driven by a superimposed dc and ac force. Shapiro steps appear at normalized voltages $(\Phi_0 f n)$ for microwave-irradiated superconducting weak links (Josephson junctions) \cite{Sha63prl} and at $(\Phi_0 f N n)$ for the motion of an Abrikosov vortex lattice under the action of superimposed dc and rf currents \cite{Fio71prl}. Here, $n$ is an integer, $f$ is the frequency of the ac stimulus, and $N$ is the number of vortex rows between the voltage leads.

We believe that in our system, magnon Shapiro steps occur because of the synchronization of the moving vortex lattice with the spin wave it excites. Specifically, the excited spin wave interacts with the vortex lattice via eddy currents induced in the superconducting strip. The presence of eddy currents is revealed via the enhancement of the instability velocity to $v^\ast_\mathrm{Ch}$ which exceeds the instability velocity in the reference state, $v^\ast$, by about 10\% (Fig.\,\ref{f3}\,\textbf{e}). An enhancement of $v^\ast$ in the superconducting strip is seen even when there is no magnon Shapiro step in the $I$-$V$ curve (compare Fig. \ref{f2}\,\textbf{a} and \ref{f2}\,\textbf{b}). Our following analysis of the vortex-lattice structure by numerical simulations suggests that the found effect is connected with preventing of the formation of ``vortex rivers''\,\cite{Vod19sst} by the eddy currents. Such ``vortex rivers'', which are self-organized Josephson-like junctions, are dynamically formed regions with a suppressed superconducting order parameter, which appear as precursors of the flux-flow instability\,\cite{Vod19sst}.
\begin{figure*}[t!]
    \centering
    \includegraphics[width=0.9\linewidth]{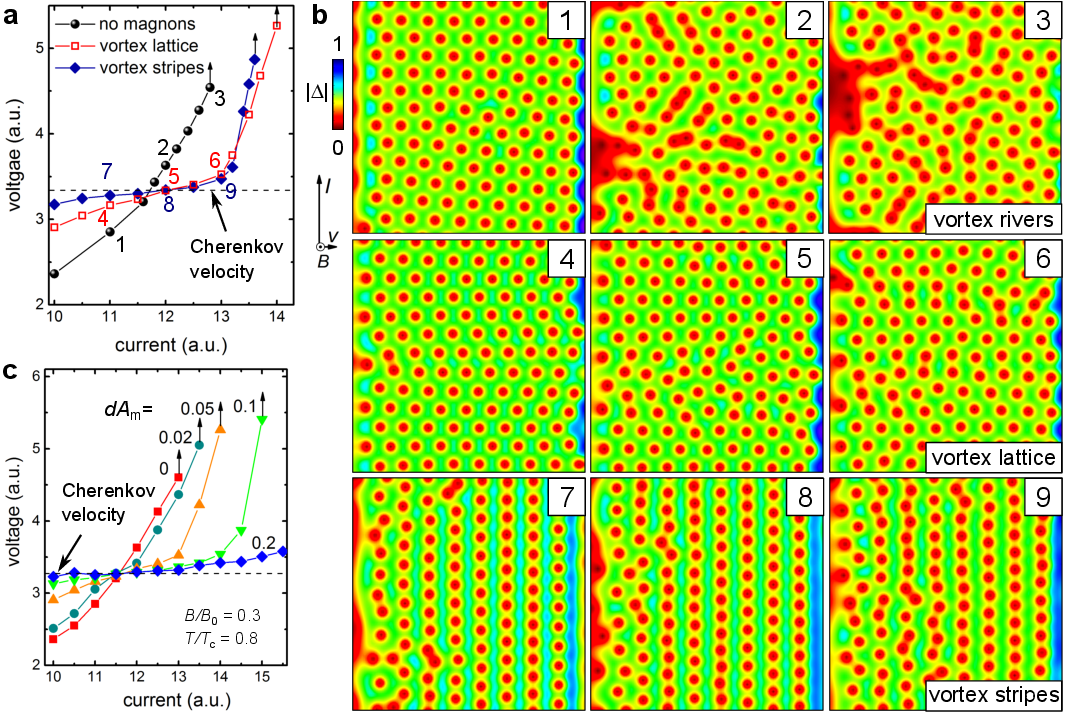}
    \caption{\textbf{TDGL simulations of the vortex motion upon Cherenkov radiation of magnons}.
    \textbf{a} $I$-$V$ curves for a superconducting strip without magnon radiation and for two cases of Cherenkov radiation of magnons accompanied by a reconfiguration of the vortex lattice.
    \textbf{b} Snapshots of the superconducting order parameter $|\Delta|$ at points 1-9 in the $I$-$V$ curves in panel (a).
    \textbf{c} Development of the magnon Shapiro step in the $I$-$V$ curve of the superconductor with increase of the amplitude $dA_\mathrm{m}$ of eddy currents upon Cherenkov radiation of spin waves.
    \label{f5}
    }
\end{figure*}

\textbf{Elucidation of the Cherenkov radiation of magnons by a lattice of fluxons}.
To analyze the excitation of spin waves by the moving vortex lattice, micromagnetic simulations were performed using the MuMax3 solver\,\cite{Van14aip}, as detailed in the supplementary materials. In the considered geometry, the spin-wave dispersion relation can be approximated by the quadratic law $f_\mathrm{SW}(k_\mathrm{SW}) = f_\mathrm{FMR} + \tilde f_\mathrm{SW}(k_\mathrm{SW})$ with $\tilde f_\mathrm{SW}(k_\mathrm{SW}) \thicksim A k_\mathrm{SW}^2$ (Fig.\,\ref{f4}\,\textbf{a}). Here, the ferromagnetic resonance frequency $f_\mathrm{FMR}$ determines the minimal frequency at $k_\mathrm{SW}=0$ and $A$ is the exchange constant. The dispersion for the moving vortex lattice is linear, with $f_\mathrm{VL}= v_\mathrm{VL}/a_\mathrm{VL} =  v_\mathrm{VL}k_\mathrm{VL}/2\pi$\,\cite{Bes14prb}. Accordingly, an increase of the vortex velocity $v_\mathrm{VL}$ leads to a steeper slope of the straight line $f_\mathrm{VL}(k_\mathrm{VL})$, which eventually intersects the parabola (Fig.\,\ref{f4}\,\textbf{a}). The condition for the dispersion crossing is $f_\mathrm{SW}= f_\mathrm{VL}$ and $k_\mathrm{SW}= k_\mathrm{VL}$. This resonance-like condition is the Cherenkov radiation condition (Magnonic regime I). With a further increase of $v_\mathrm{VL}$, the straight line $f_\mathrm{VL}(k_\mathrm{VL})$ intersects the parabola at two points. A closer insight into the physics of Cherenkov radiation of magnons by a lattice of fluxons can be gained from a consideration of the dependences $v_\mathrm{SW}(k_\mathrm{SW})$ deduced from the dispersion curves.

Figure \ref{f4}\,\textbf{b} presents the dependence of the spin-wave phase and group velocities, $v_\mathrm{ph} = 2\pi f_\mathrm{SW}/k_\mathrm{SW}$ and $v_\mathrm{g} = 2\pi \partial f_\mathrm{SW}/\partial k_\mathrm{SW}$, on the wavenumber $k_\mathrm{SW}$. In this representation, different vortex lattice velocities correspond to the crossings of $v_\mathrm{ph}(k_\mathrm{SW})$ and $v_\mathrm{g}(k_\mathrm{SW})$ at different $v$ levels parallel to the $k$-axis. At the Cherenkov resonance condition, not only $v_\mathrm{VL}$ is equal to $v_\mathrm{g}$, but also $v_\mathrm{g} = v_\mathrm{ph}$. This means that the energy from the vortex lattice is spent for a monochromatic excitation of spin waves which, in addition, are excited unidirectionally, i.e. propagate only in one direction which is determined by the direction of motion of the vortex lattice. The excited spin wave propagates towards the detector region where it is efficiently collected by the specially designed detector antenna, resulting in a well-defined frequency peak (Fig.\,\ref{f4}\,\textbf{c}).

Distinct from this, out of the wavenumber resonant condition, when $v_\mathrm{VL}$ exceeds the threshold velocity of the Cherenkov radiation, two spin waves are excited with different group velocities and wavelengths (see the interference pattern in the top panel of Fig. 4\,\textbf{d}). In this regime, which we call the Magnonic regime II, one spin wave moves faster than the vortex lattice and the other one moves slower. However, due to the very weak excitation and detection efficiency for these wavelengths, their intensities are negligibly small (Fig.\,\ref{f4}\,\textbf{c}). The excitation and propagation of spin waves in the different dynamic regimes is illustrated further in the supplementary materials (supplementary text and movies 1-3). The estimated attenuation length of the generated spin waves (at $k_\mathrm{SW}\approx175\,\mu$m) is around 600\,nm \cite{Kal86jpc}. Together with sub-40\,nm wavelengths this makes a fast-moving Abrikosov vortex lattice an interesting source for spin-wave excitation in cryogenic magnonics.

\textbf{Evolution of vortex lattice configurations upon Cherenkov radiation of magnons}.
For analysis of the vortex lattice configurations at the Cherenkov radiation of magnons, simulations relying upon the solution of a modified time-dependent Ginzburg-Landau (TDGL) equation have been performed. At the Cherenkov resonance condition found from the micromagnetic simulations (Fig.\,\ref{f4}\,\textbf{a}), spin-wave-induced eddy currents in the superconductor were phenomenologically introduced by the term $\mathbf{A}_\mathrm{m}$ added to the vector potential $\mathbf{A}$ in the TDGL. Details on the TDGL simulations are provided in the supplementary materials.

The simulated $I$-$V$ curves for the bare superconducting strip and the superconductor/ferro\-magnet heterostructure are presented in Fig.\,\ref{f5}\,\textbf{a}. With increase of the amplitude of the eddy currents, a constant-voltage step is developed in the $I$-$V$ curve (Fig.\,\ref{f5}\,\textbf{c}). This is accompanied by a dynamic ordering of the vortex lattice, as concluded from the snapshots of the superconducting order parameter $|\Delta|$ (Fig.\,\ref{f5}\,\textbf{b}). These snapshots a shown for a selection of points in the $I$-$V$ curves calculated without magnon radiation and with the magnon radiation for two symmetries of the eddy currents (``vortex lattice'' and ``vortex stripes'' in Fig.\,\ref{f5}\,\textbf{a}). Specifically, without magnon radiation, the flow of vortices ordered in a nearly hexagonal vortex lattice (snapshot 1) loses its long-range order (snapshot 2) in the regime of nonlinear conductivity II (see also Fig.\,\ref{f2}\,\textbf{a}). As the transport current increases, areas with a suppressed order parameter are nucleated at the edge where the vortices enter the superconductor, leading to the formation of ``vortex rivers'' (snapshot 3). Within the framework of the theory of edge-controlled flux-flow instability \cite{Vod19sst}, the development of vortex rivers leads to the formation of normally conducting domains across the superconducting strip. This results in an avalanche-like transition of the sample to the normal state.

When the excitation of magnons is modeled with a resonance-like enhancement of the eddy currents, the $I$-$V$ curve becomes flattened and the motion of vortices ordered in a hexagonal lattice (snapshots 4, 5 and 6) or stripes (snapshots 7, 8 and 9) persists up to larger currents. Note, that in the case of vortex stripes the section in the $I$-$V$ curve corresponding to the Cherenkov velocity reproduces the magnon Shapiro steps observed experimentally. With a further current increase, the vortex stripes hinder the development of vortex rivers such that an ordered vortex-stripe state is preserved up to yet larger transport currents (snapshots 8 and 9) as compared to the  case of hexagonal vortex lattice. These results are valid only when the symmetry of the moving vortex lattice corresponds to the symmetry of the eddy currents. For example, at point 6 the ordering of vortices deviates strongly from the assumed nearly triangular vortex lattice (snapshot 6) and a transition to a non-ordered state occurs. Separately for the stripe symmetry, the vortex lattice maintains its stripe-like ordering in a broad range of transport currents despite the same amplitude of the eddy currents.

In the experiment, the coupling strength between the fluxon and magnon subsystems (i.e. the amplitude and spatial distribution of the eddy currents) is a poorly accessible quantity. In the simulations, the introduction of the amplitude of the eddy currents as a free parameter allows us to demonstrate a continuous evolution of the $I$-$V$ curves from the nonlinear conductivity regime II followed by the instability jump regime III (as for the bare superconductor reference sample, Fig.\,\ref{f1}\,\textbf{a}) to the voltage step regime M1 followed by the steep upturn regime M2 and the instability jump regime III in the $I$-$V$ curve for the superconductor/ferromagnet heterostructure (Fig.\,\ref{f5}\,\textbf{c}). In addition, the TDGL simulations illustrate that a higher instability current $I^\ast_\mathrm{Ch}$ and instability velocity $v^\ast_\mathrm{Ch}$ can indeed be achieved upon the Cherenkov radiation of magnons by fast-moving fluxons. The enhancement of $I^\ast_\mathrm{Ch}$ and $v^\ast_\mathrm{Ch}$ occurs because the transition of the moving vortex lattice to the vortex river regime is prevented by the eddy currents induced by the excited spin wave for both the vortex stripe and the vortex lattice symmetries.

Finally, regarding the sub-40\,nm wavelengths (wavenumbers $k_\mathrm{SW}>170$\,rad/$\mu$m) of the excited/detected spin waves, we should emphasize that the excitation of propagating spin waves with wavelengths below 100\,nm represents a critical task of modern magnonics\,\cite{Yuh16nac,Liu18nac,Slu19nan,Die19prl,Che20nac,Yuh21phr}. Excitation of short-wavelength exchange spin waves is challenging because of low microwave-to-magnon conversion efficiencies at the sub-100 nm scale. Here, we have demonstrated a new paradigm for the excitation of short-wave (exchange) spin waves by Abrikosov vortices as fast-moving magnetic perturbations, with wavelengths by about a factor of two smaller than the shortest wavelengths of spin waves previously observed experimentally\,\cite{Yuh16nac,Liu18nac,Slu19nan,Die19prl,Che20nac,Yuh21phr}. In addition, because of the periodicity of the vortex lattice, the spin-wave excitation is unidirectional (spin wave propagates in the direction of the vortex motion) and monochromatic (wavelength is equal to the vortex lattice parameter). This makes the magnon Cherenkov radiation by fast-moving vortices an excellent spin-wave source for magnonic applications at the nano-scale.

Summarizing, Cherenkov radiation of magnons by fast-moving fluxons has been experimentally realized in a ferromagnet/superconductor heterostructure. The phenomenon has been evidenced by the magnon Shapiro step in the current-voltage curve of the superconductor and by the direct microwave detection. It is found that the frequency of the excited spin wave is equal to the washboard frequency of the vortex lattice while its wavelength is determined by the vortex lattice period. The excitation is monochromatic and unidirectional due to the match of the spin-wave group and phase velocities with the velocity of the vortex lattice. In combination with the excitation of very short (sub-40\,nm) wavelengths, inaccessible by other approaches, this makes the magnon Cherenkov radiation by fast-moving fluxons a valuable spin-wave generation mechanism for applied magnonics. Finally, we have demonstrated that the magnon radiation preserves the long-range order of the vortex lattice and enhances the current-carrying ability of the superconductor.
\vspace{3mm}

\begin{acknowledgments}
Precursors were provided courtesy of Sven Barth (Goethe University Frankfurt). O.V.D. is very grateful to Menachem Tsindlekht (Hebrew University of Jerusalem) for support with the microwave detection experiments.
The authors are grateful to Andrii Chumak for fruitful discussions. 
O.V.D. acknowledges the German Research Foundation (DFG) for support through Grant No. 374052683 (DO1511/3-1).
B.B. acknowledges financial support by the Vienna Doctoral School in Physics (VDSP).
Q.W. acknowledges support within the ERC Starting Grant no. 678309 MagnonCircuits.
Support through the Frankfurt Center of Electron Microscopy (FCEM) and by the European Cooperation in Science and Technology via COST Action CA16218 (NANOCOHYBRI) is gratefully acknowledged.
\end{acknowledgments}

{\bf Author contributions:}
O.V.D. conceived the experiment and performed the measurements.
Q.W. performed micromagnetic simulations.
D.Y.V. performed TDGL simulations.
B.B. evaluated the data.
R.S. fabricated the samples and automated the data acquisition.
A.I.B. provided theoretical support.
O.V.D., D.Y.V., S.K., M.H. and A.I.B. discussed the interpretation and the relevance of the results.
O.V.D. wrote the first version of the manuscript.
All authors discussed the results and contributed to the manuscript writing.





\section*{Materials and Methods}

\textbf{Fabrication of the microwave nano-antenna}. Fabrication of the system began with the deposition of a 40/5\,nm Au/Cr film onto a Si\,(100,\,p-doped)/SiO$_2$\,(200\,nm) substrate and its patterning for electrical dc and microwave measurements. In the sputtering process, the substrate temperature was $T=22^\circ$C, the growth rates were 0.055\,nm/s and 0.25\,nm/s, and the Ar pressures were 2$\times$10$^{-3}$\,mbar and 7$\times$10$^{-3}$\,mbar for the Cr an Au layers, respectively. The microwave ladder antenna was fabricated from the Au/Cr film by focused ion beam (FIB) milling at 30\,kV/30\,pA in a dual-beam scanning electron microscope (FEI Nova Nanolab 600). The multi-element antenna consisted of ten nanowires connected in parallel between the signal and ground leads of a 50\,$\Omega$-matched microwave transmission line. The antenna had a period $p =108$\,nm with the nanowire width equal to the nanowire spacing, so that its Fourier transform contained only odd spatial harmonics with $\lambda_1 = p$ and $\lambda_3 = \lambda_1/3 = 36$\,nm, that made it sensitive to spin-wave wavelengths of 36$\pm$2\,nm in our experiments.

\textbf{Fabrication of the Nb-C superconducting microstrip}. Fabrication of the ladder antenna was followed by direct-writing of the superconducting strip, at a distance of 2\,$\mu$m away (edge-to-edge) from the microwave antenna. The 45\,nm-thick Nb-C microstrip was fabricated by focused ion beam induced deposition (FIBID). FIBID was done at 30\,kV/10\,pA, 30\,nm pitch and 200\,ns dwell time employing Nb(NMe$_2$)$_3$(N-$\textit{t}$-Bu) as precursor gas\,\cite{Por19acs}. The superconducting strip and the ladder antenna were covered with a 3\,nm-thick insulating Nb-C layer prepared by focused electron beam induced deposition (FEBID). Before the deposition of the Co-Fe magnonic waveguide a 48\,nm-thick insulating Nb-C-FEBID layer was deposited to compensate for the structure height variations between the antenna and the Nb-C strip. The elemental composition in the Nb-C microstrip is 45$\pm$2\,\%\,at.\,C, 29$\pm$2\,\%\,at.\,Nb, 15$\pm$2\,\%\,at.\,Ga, and 13$\pm$2\,\%\,at.\,N, as inferred from energy-dispersive X-ray spectroscopy on thicker microstrips written with the same deposition parameters. The Nb-C microstrip had well-defined smooth edges and an rms surface roughness of <0.3\,nm, as deduced from atomic force microscopy scans over its 1\,$\mu$m$\times$1\,$\mu$m active part, prior to the deposition of the Co-Fe layer. The distance between the voltage leads was 1\,$\mu$m. To prevent current-crowding effects at the sharp strip edges, which may lead to an undesirable reduction of the experimentally measured critical current (and the instability current), the two ends of the strip had rounded sections \cite{Kor20pra}.

\textbf{Superconducting properties of the Nb-C microstrip}. The resistivity of the microstrip at 7\,K was $\rho_\mathrm{7K} = 551\,\mu\Omega$cm. Below the transition temperature $T_\mathrm{c}=5.60$\,K, deduced by using a 50\% resistance drop criterion, the microstrip transited to a superconducting state. Application of a magnetic field $H_\mathrm{ext}\approx2$\,T led to a decrease of $T_\mathrm{c}(0)$ to $T_\mathrm{c}(\mathrm{2\,T})\approx5.1$\,K. Near $T_\mathrm{c}$, the critical field slope $dH_{\mathrm{c}2}/dT|_{T_\mathrm{c}} = -2.19$\,T\,K$^{-1}$ corresponds, in the dirty superconductor, to the electron diffusivity $D = -4 k_\mathrm{H}/[\pi e (dH_{\mathrm{c}2}/dT|_{T_\mathrm{c}})] \approx 0.5$\,cm$^2$\,s$^{-1}$ with the extrapolated zero-temperature upper critical field value $H_\mathrm{c2}(0)=12.3$\,T. The coherence length and the penetration depth at zero temperature were estimated\,\cite{Kor18pra} as $\xi_c = \sqrt{\hbar D /k_\mathrm{B}T_\mathrm{c}} = 9$\,nm (corresponding to $\xi(0) = \xi_\mathrm{c}/\sqrt{1.76} = 7$\,nm) and $\lambda(0) = 1.05\cdot10^{-3} \sqrt{\rho_\mathrm{7K} k_\mathrm{B}/\Delta(0)} \approx 1040\,$nm. Here, $\Delta(0)$ is the zero-temperature superconducting energy gap and $\hbar$ the Planck constant.

\textbf{Fabrication and properties of the Co-Fe magnonic conduit}. The Co-Fe microstrip was 1\,$\mu$m wide, 5\,$\mu$m long and 30\,nm thick. It was fabricated by FEBID employing HCo$_3$Fe(CO)$_{12}$ as precursor gas\,\cite{Por15nan,Kum18jpc,Dob19ami}. FEBID was done with 5\,kV/1.6\,nA, 20\,nm pitch, and 1\,$\mu$s dwell time. The material composition in the magnonic waveguide is $61\pm3$\,at.\% Co, $20\pm3$\,at.\% Fe, $11\pm3$\,at.\% C, $8\pm3$\,at.\% C. The oxygen and carbon are residues from the precursor in the FEBID process \cite{Hut18mee}. The Co-Fe conduit consisted of a dominating bcc Co$_3$Fe phase mixed with a minor amount of FeCo$_2$O$_4$ spinel oxide phase with nanograins of about 5\,nm diameter\,\cite{Por15nan}. The random orientation of Co-Fe grains in the carbonaceous matrix ensured negligible magnetocrystalline anisotropy \cite{Por15nan}. An external field $H_\mathrm{ext}$ of about 1.75\,T was enough to magnetize the Co-Fe magnonic conduit in the direction perpendicular to its plane. Further details on the microstructural and magneto-transport properties of Co-Fe-FEBID were reported previously \cite{Por15nan}.

\textbf{Electrical resistance measurements}. The $I$-$V$ curves were acquired in the current-driven mode in a He$^4$ cryostat equipped with a superconducting solenoid. The external magnetic field $H_\mathrm{ext}$ was tilted at a small angle $\beta=5^\circ$ with respect to the normal to the sample plane ($z$ axis) in the plane perpendicular to the direction of the transport current.  The field value was varied between 1.75\,T and 1.95\,T, inducing a vortex lattice with parameter $a_\mathrm{VL} = \sqrt{2\Phi_0/\sqrt{3}H_\mathrm{ext}}$. Here, the small field tilt angle $\beta$ ensures that the field component $H_\mathrm{ext\,z}$ acting along the $z$ axis is only negligibly smaller than $H_\mathrm{ext}$ with $(H_\mathrm{ext} - H_\mathrm{ext\,z})/H_\mathrm{ext} \lesssim 0.5\%$. The transport current applied along the $y$ axis in a magnetic field $H \approx H_\mathrm{ext} = H_\mathrm{z}$ exerted on vortices a Lorentz force acting along the $x$ axis \cite{Bra95rpp}. The voltage induced by the vortex motion across the superconducting microstrip was measured with a nanovoltmeter. A series of reference measurements was taken before the deposition of the Co-Fe conduit on top of the Nb-C microstrip. No voltage steps were revealed in the $I$-$V$ curves of the bare Nb-C strip. By contrast, constant-voltage steps in the $I$-$V$ curves were revealed after the deposition of the Co-Fe magnonic conduit on top of the Nb-C strip. The vortex velocity $v_\mathrm{VL}$ was deduced from the $I$-$V$ curves by using the standard formula $v_\mathrm{VL} = V/(BL)$ \cite{Bra95rpp}, where $V$ is the measured voltage, $B$ is the induction of the external magnetic field and $L = 1\,\mu$m the distance between the voltage leads. The rarely achieved combination or properties in the Nb-C superconductor--weak volume pinning, close-to-perfect edge barrier and a fast relaxation of non-equilibrium electrons--allow for ultra-fast motion of Abrikosov vortices, which are usually not achievable in superconductors because of the onset of the flux-flow instability \cite{Sil12njp,Gri15prb,Att12pcm,Cap17apl,Leo11prb}.

\textbf{Microwave detection of spin waves}. The microwave detection of spin waves was done using a microwave ladder nano-antenna connected to a spectrometer system which allowed for the detection of signals at power levels down to 10$^{-16}$\,W in a 25\,MHz bandwidth \cite{Dob18nac}. The detector system consisted of a spectrum analyzer (Keysight Technologies N9020B, 10\,Hz-50\,GHz), a semirigid coaxial cable (SS304/BeCu, dc-61 GHz), and an ultra-wide-band low-noise amplifier (RF-Lambda RLNA00M54GA, 0.05-54\,GHz).

\textbf{Micromagnetic simulations}. The micromagnetic simulations were performed by the GPU-accelerated simulation package MuMax3 to calculate the space- and time-dependent magnetization dynamics in the investigated structures\,\cite{Van14aip}. In the simulations, following parameters were used for the Co-Fe magnonic conduit: saturation magnetization $M_\mathrm{s} = 1.4-1.5\times$\,MA/m, exchange constant $A = 15-18$\,pJ/m, and Gilbert damping $\alpha = 0.01$. The best match of the simulation results with the experimental data has been revealed for $M_\mathrm{s} = 1.45$\,MA/m and $A = 17$\,pJ/m. The mesh was set to $2\times2$\,nm$^2$, which is smaller than the exchange length of Co-Fe ($\approx5$\,nm). An external field $H_\mathrm{ext}$ in the range 1.75-1.95\,T, which was sufficient to magnetize the structure to saturation in the out-of-plane direction, was applied at a small angle $\beta$ with respect to the $z$ axis in the $xz$ plane. A fast-moving periodic field modulation was used to mimic the effect of the moving vortex lattice. The oscillations $m_\mathrm{x}(x,y,t)$ were calculated for all cells and all times via $m_\mathrm{x}(x,y,t)=M_\mathrm{x}(x,y,t) - M_\mathrm{x}(x,y,0)$, where $M_\mathrm{x}(x,y,0)$ corresponds to the ground state (fully relaxed state without any moving magnetic field source). The dispersion curves were obtained by performing two-dimensional fast Fourier transformations of the time- and space-dependent data. The spin-wave spectra were calculated by performing a fast Fourier transformation of the data in a region which was at a distance of $1\,\mu$m away from the spin-wave excitation region.

The simulation results were first validated by a comparison with the results of analytical calculations. Namely, the spin-wave dispersion curve for the Co-Fe was first compared with the dispersion curve calculated within the framework of the Kalinikos-Slavin theory\,\cite{Kal86jpc}. It should be noted that the relevant angle in the Kalinikos-Slavin theory is the angle $\theta$ at which the effective magnetic field $H_\mathrm{eff}$ is tilted with respect to the $z$ axis. This angle $\theta$ and the effective field $H_\mathrm{eff}$ were extracted from micromagnetic simulations. Boundary conditions of fully pinned spins at the edges of the Co-Fe conduit were used. The dispersion curves calculated by using the analytical Kalinikos-Slavin theory fit very well with the simulation results.

In the investigated range of fields about 2\,T, the vortex lattice is dense and the modulation of the local magnetic field along $z$-component at the vortex cores and between them is small, with $\Delta B_\mathrm{z} \approx 0.1$\,mT. This is because the magnetic penetration depth $\lambda$ ($\sim1\,\mu$m) is much larger than the vortex-lattice parameter $a_\mathrm{VL}$ ($\approx35$\,nm) in the Nb-C strip. However, the other components of the field modulation also contribute to the spin-wave excitation. Given the large number of vortices (850-950, depending on the applied field value) threading the $1\mu$m$\times1\,\mu$m part of the superconductor underneath the Co-Fe magnonic conduit at the magnetic fields of interest, such a small modulation of the magnetic field induced by the moving vortex lattice was enough to excite spin waves propagating over the $2\,\mu$m distance between the superconducting microstrip and the microwave antenna. Within the framework of the Kalinikos-Slavin theory \cite{Kal86jpc}, the spin-wave decay length was estimated as $600$\,nm at the wavenumber $k_\mathrm{SW} = 175$\,rad/$\mu$m.

In the experiment, the external field $\mathbf{H}_\mathrm{ext}$ was directed at a small tilt angle $\beta$ with respect to the normal to the sample plane. The angle $\beta$ lied in the $xz$-plane and it was nominally set to $\beta = 5^\circ$ in the experiment. It should be noted that distinct from the limiting case of forward volume spin-wave (FVSW) geometry with $\beta = 0^\circ$, the magnetization $\mathbf{M}$ of the Co-Fe conduit at $\beta \neq 0^\circ$ is directed not along $\mathbf{H}_\mathrm{ext}$, but along the effective field $\mathbf{H}_\mathrm{eff}$ tilted at the angle $\theta$ away from the $z$-axis. The angle $\theta$ and the effective field $H_\mathrm{eff}$ depend strongly on the angle $\beta$. The dependences $\theta(\beta)$ were deduced from the micromagnetic simulations for a series of values of the saturation magnetization $M_\mathrm{s}$, exchange stiffness $A$, and the thickness of the Co-Fe waveguide.

Various spatial field profiles and arrangements of vortices as moving magnetic perturbations were used in the simulations. Namely, the excitation of spin waves was checked for saw-tooth, cosine and meander-like magnetic induction profiles, as well as for vortices ordered in a hexagonal, square and squeezed-square (stripe-like pattern) lattices. The largest spin-wave amplitude was achieved with a field modulation induced by a moving array of periodically arranged vortex stripes, while the smallest spin-wave amplitude resulted for a hexagonal vortex lattice.

\textbf{Time-dependent Ginzburg-Landau simulations}. The evolution of the superconducting order parameter $\Delta=|\Delta|e^{i\phi}$ was analyzed relying upon a numerical solution of the modified TDGL equation\,\cite{Vod17pra}
\begin{eqnarray*}
    \frac{\pi\hbar}{8k_\mathrm{B}T_\mathrm{c}} \left(\frac{\partial }{\partial t}+\frac{2ie\varphi}{\hbar} \right) \Delta= \nonumber
    \\
    =\xi^2_\mathrm{mod}\left( \nabla -i\frac{2e}{\hbar c}A\right)^2\Delta+\left(1-\frac{T_\mathrm{e}}{T_\mathrm{c}}-\frac{|\Delta|^2}{\Delta_{mod}^2}\right)\Delta+
    \nonumber
    \\
    +i\frac{(\mathrm{div}\mathbf{j}_\mathrm{s}^\mathrm{Us}-\mathrm{div}\mathbf{j}_\mathrm{s}^\mathrm{GL})}{|\Delta|^2}\frac{e\Delta\hbar D}{\sigma_\mathrm{n}\sqrt{2}\sqrt{1+T_\mathrm{e}/T_\mathrm{c}}},
\end{eqnarray*}
where $\xi^2_\mathrm{mod}=\pi\sqrt{2}\hbar D/(8k_\mathrm{B}T_\mathrm{c}\sqrt{1+T_\mathrm{e}/T_\mathrm{c}})$,
$\Delta_\mathrm{mod}^2=(\Delta_0\tanh(1.74\sqrt{T_\mathrm{c}/T_\mathrm{e}-1}))^2/(1-T_\mathrm{e}/T_\mathrm{c})$,
$A$ is the vector potential,
$\varphi$ is the electrostatic potential,
$D$ is the diffusion coefficient,
$\sigma_\mathrm{n}=2e^2DN(0)$ is the normal-state conductivity with $N(0)$ being the single-spin density of states at the Fermi level,
$T_\mathrm{e}$ and $T_\mathrm{p}$ are the electron and phonon temperatures,
and
$\mathbf{j}_\mathrm{s}^\mathrm{Us}$ and
$\mathbf{j}_s^\mathrm{GL}$ are the superconducting current densities in the Usadel and Ginzburg-Landau models
\begin{eqnarray}
    \mathbf{j}_\mathrm{s}^{\mathrm{Us}}=\frac{\pi\sigma_\mathrm{n}}{2e\hbar}|\Delta|\tanh
    \left(\frac{|\Delta|}{2k_\mathrm{B}T_\mathrm{e}}\right)\mathbf{q}_\mathrm{s},
\end{eqnarray}
where $\mathbf{q}_s=\nabla \varphi+2\pi A/\Phi_0$, and
$\mathbf{j}_\mathrm{s}^{\mathrm{GL}}=\displaystyle\frac{\pi\sigma_\mathrm{n}|\Delta|^2}{4ek_\mathrm{B}T_\mathrm{c}
\hbar}\mathbf{q}_\mathrm{s}$.

The vector potential ${\bf A}=(0,A_\mathrm{y},0)$ in the TDGL equation consists of two parts: $A_\mathrm{y}=H_\mathrm{ext}x+A_\mathrm{m}$, where $H_\mathrm{ext}$ is the external magnetic field and $A_\mathrm{m}$ is the vector potential of the magnetic field induced in the superconducting strip by spin waves. Two kinds of $A_\mathrm{m}$ are considered
\begin{eqnarray*}
    A_\mathrm{m} = dA_\mathrm{m} \sin(2\pi(y-v_\mathrm{Ch}t)/a_\mathrm{x})\sin(2\pi x/a_\mathrm{y}),
    \\
    A_\mathrm{m} = dA_\mathrm{m} \sin(2\pi(y-v_\mathrm{Ch}t)/a_\mathrm{x}),
\end{eqnarray*}
where $v_\mathrm{Ch}$ is the Cherenkov velocity and $a_\mathrm{x}$ and $a_\mathrm{y}$ are parameters
of the order of $a_\mathrm{VL}$. The first expression for $A_\mathrm{m}$ models the resonance response of the
ferromagnet in the presence of a nearly triangular vortex lattice moving with the velocity $v_\mathrm{Ch}$. The second expression for $A_\mathrm{m}$ accounts for the assumed appearance of vortex stripes at the resonance condition. Physically, the second expression is connected with the much larger amplitude of spin waves (inducing a larger $dA$) in comparison with spin waves excited by a triangular vortex lattice, as inferred from the micromagnetic simulations. The component of the vector potential $A_\mathrm{m}$ induces eddy
supercurrents in the superconductor, which affect the vortex motion. The amplitude $dA_\mathrm{m}$ controls the amplitude of the eddy currents in the considered model as the relation between the superconducting
eddy currents and $A_\mathrm{m}$ follows from the equation for $\mathbf{j}_\mathrm{s}^\mathrm{Us}$.

The electron and phonon temperatures, $T_\mathrm{e}$ and $T_\mathrm{p}$, were found from the solution of following equations
\begin{eqnarray*}
    \frac{\partial}{\partial t}\left(\frac{\pi^2k_\mathrm{B}^2N(0)T_\mathrm{e}^2}{3}-
    \mathcal{E}_0\mathcal{E}_\mathrm{s}(T_\mathrm{e},|\Delta|)\right) = \nonumber
        \\
    = \nabla k_\mathrm{s} \nabla T_\mathrm{e}-\frac{96\zeta(5)N(0)k_\mathrm{B}^2}{\tau_0}\frac{T_\mathrm{e}^5-
    T_\mathrm{p}^5}{T_\mathrm{c}^3}+ j E,
        \\
    \frac{\partial T_\mathrm{p}^4}{\partial t}=-\frac{T_\mathrm{p}^4- T^4}{\tau_\mathrm{esc}}+\gamma\frac{24\zeta(5)}{\tau_0}\frac{15}{\pi^4}\frac{T_\mathrm{e}^5-T_\mathrm{p}^5}{T_\mathrm{c}},
\end{eqnarray*}
where $\mathcal{E}_0=4N(0)(k_\mathrm{B}T_\mathrm{c})^2$, $\mathcal{E}_0\mathcal{E}_\mathrm{s}(T_\mathrm{e},|\Delta|)$ is the change in the energy of electrons due to the transition to the superconducting state, $k_\mathrm{s}$ is the heat conductivity in the superconducting state
\begin{equation*}
    k_\mathrm{s}=k_\mathrm{n}\left(1-\frac{6}{\pi^2(k_\mathrm{B}T_\mathrm{e})^3}\int_0^{|\Delta|}\frac{\epsilon^2
    e^{\epsilon/k_\mathrm{B}T_\mathrm{e}}d\epsilon}{(e^{\epsilon/k_\mathrm{B}T_\mathrm{e}}+1)^2}\right),
\end{equation*}
$k_\mathrm{n} = 2D\pi^2k_\mathrm{B}^2N(0)T_\mathrm{e}/3$ is the heat conductivity in the normal state, the term $jE$ describes Joule dissipation, and $\tau_\mathrm{esc}$ is the escape time of nonequilibrium phonons to the substrate. The parameter $\gamma$ is defined as $\gamma= \displaystyle\frac{8\pi^2}{5}\displaystyle\frac{C_\mathrm{e}(T_\mathrm{c})}{C_\mathrm{p}(T_\mathrm{c})}$, where $C_\mathrm{e}(T_\mathrm{c})$ and $C_\mathrm{p}(T_\mathrm{c})$ are the heat capacities of electrons and phonons at $T=T_\mathrm{c}$, and the characteristic time $\tau_0$ controls the strength of the electron-phonon and phonon-electron scattering\,\cite{Vod17pra}.

The electrostatic potential $\varphi$ was found from the current continuity equation
\begin{equation*}
    \mathrm{div}(\mathbf{j}_\mathrm{s}^{Us}+\mathbf{j}_\mathrm{n})=0,
\end{equation*}
where $\mathbf{j}_\mathrm{n}=-\sigma_\mathrm{n}\nabla \varphi$ is the normal current density.

The boundary conditions at the microstrip edges, where vortices enter and exit it, were $\mathbf{j}_\mathrm{n}|_\mathrm{n}=\mathbf{j}_\mathrm{s}|_\mathrm{n}=0$ and $\partial T_\mathrm{e}/\partial\mathrm{n}=0$, $\partial |\Delta|/\partial n=0$. At the edges along the current direction the boundary conditions were $T_\mathrm{e}=T$, $|\Delta|=0$, $\mathbf{j}_\mathrm{s}|_\mathrm{n}=0$, and $\mathbf{j}_\mathrm{n}|_\mathrm{n}=I/wd$. The latter boundary conditions model the contact of the superconducting strip with a normal reservoir being in equilibrium. This choice provides a way to inject the current into the superconducting microstrip in the modeling.

For the simulation results shown in Fig.\,\ref{f5}, the modeled length of the microstrip is $L=4w$, the width $w = 50\xi_\mathrm{c}$, the parameter $B_0 = \Phi_0/(2\pi\xi_\mathrm{c}^2) \simeq 4.15$\,T, where $\xi_\mathrm{c}= 8.9$\,nm. The calculations were done with parameters $\gamma=9$ and $\tau_0=925$\,ns for NbN as their values for NbC are unknown, but supposed to be of the same order of magnitude. A variation of $\gamma$ and $\tau$ only leads to quantitative changes in the $I$-$V$ curves and without qualitative changes in the vortex dynamics. In simulations $dA$ was varied between 0 (no ferromagnet layer) and $0.1\Phi_0/2\pi\xi_\mathrm{c}$ which corresponds to about of 1/4 of the depairing velocity for superconducting charge carriers (Cooper pairs) or critical $q_\mathrm{s}^\mathrm{c} \sim 0.35\Phi_0/2\pi\xi_\mathrm{c}$ of the superconducting strip at $B=0$ and $T=0.8 T_\mathrm{c}$. The parameters $a_\mathrm{x}$ and $a_\mathrm{y}$ were chosen to model a triangular moving vortex lattice in absence of the ferromagnetic layer and far from the instability point (see snapshot 4 for the distribution of the superconducting order parameter in Fig.\,\ref{f5}(b)). In Fig.\,\ref{f5} we present the results for $v=110\xi_\mathrm{c}/\tau_0$, $a_\mathrm{x}=5.5 \xi_\mathrm{c}$ and $a_\mathrm{y}=9.2 \xi_\mathrm{c}$ ($a_\mathrm{VL}=4.9 \xi_\mathrm{c}$ at $B=0.3B_0$). We find that the width and the slope of the  ``plateau'' in the $I$-$V$ curve weakly vary with small variations of $a_\mathrm{x}$ and $a_\mathrm{y}$, while the value of $v_\mathrm{Ch}$ controls the position of the voltage ``plateau''.


%

\end{document}